\documentstyle[12pt]{article}
\newcommand{\bq}{\begin{equation}}
\newcommand{\eq}{\end{equation}}
\newcommand{\ba}{\begin{eqnarray}}
\newcommand{\ea}{\end{eqnarray}}

\newcommand{\nl }{ \nonumber  }

\newcommand{\p}{\partial}
\newcommand{\pu}{\p_\tau}
\newcommand{\pj}{\p_j}
\newcommand{\h}{\hspace{1cm}}
\newcommand{\s}{\sigma}
\newcommand{\us}{\underline\sigma}
\newcommand{\uz}{\underline{z}}

\begin{document}
\vspace*{2cm}
{\bf\begin{center}
 N=1, D=10 TENSIONLESS SUPERBRANES II.
\footnote{Work supported in part by the National Science Foundation
of Bulgaria under contract $\phi-620/1996$}
\vspace*{1cm}
\\
P. Bozhilov
\footnote {E-mail: bojilov@thsun1.jinr.ru; permanent address:
Dept.of Theoretical Physics,"Konstantin Preslavsky" Univ. of 
Shoumen, 9700 Shoumen, Bulgaria} 
\\
\it
Bogoliubov Laboratory of Theoretical Physics, \\
JINR, 141980 Dubna, Russia \\  
\vspace*{1cm} 
\end{center}}

We consider a model for tensionless (null) $p$-branes with $N=1$ 
global supersymmetry in 10-dimensional Minkowski space-time.
We give an action for the model and show that it is 
reparametrization and kappa-invariant.
We also find some solutions of the classical equations of motion.
In the case of null superstring ($p=1$), we obtain the general 
solution in arbitrary gauge.
\vspace*{1cm}
\section{\bf Introduction}
\hspace{1cm}
The null $p$-branes are the zero tension limit of the
tensionful ones. The correspondence between this two types of 
branes may be regarded as a generalization of the massless-massive
particles relationship.
Null branes with manifest space-time or world-volume 
supersymmetry are considered in \cite{Z} and \cite{S} respectively. 
In a previous paper \cite{PLB98}, we began the investigation 
of a tensionless p-brane model with $N=1$ supersymmetry in 
ten dimensional flat space-time. Starting with a Hamiltonian which 
is a linear combination of first and mixed (first and second) class 
constraints, we succeed to obtain a new one, which is a linear 
combination of first class, BFV-irreducible and Lorentz-covariant 
constraints only. This was done with the help of the introduced 
auxiliary harmonic variables \cite{Sok}, \cite{NissP}. 
Then we gave manifest expressions for 
the classical BRST charge, the corresponding total constraints and 
BRST-invariant Hamiltonian.

In this letter, we continue the investigation of the model. Here, 
we consider the corresponding action, establish its symmetries, and 
present some solutions of the classical equations of motion.

Our initial Hamiltonian is \cite{PLB98}
\ba 
\label{H}
H_0=\int d^p\sigma\bigl [\mu^0 T_0+\mu^j T_j+\mu^\alpha 
D_{\alpha}\bigr ],
\ea
where the constraints $T_0$, $T_j$ and $D_\alpha$ are defined by 
the equalities:
\ba
\nl
T_0 &=& p_{\mu}p_{\nu}\eta^{\mu\nu},
\h
diag(\eta_{\mu\nu})=(-,+,...,+), 
\h
(\mu,\nu = 0,1,...,9),
\\
\nl
T_j &=& p_{\nu}\p_j x^{\nu}+p_{\theta\alpha}\p_j \theta^{\alpha},
\h
\p_j=\p/\p\s^j ,\h (j = 1,2,...,p),
\\
\nl
D_{\alpha}&=& -ip_{\theta\alpha}-(\not{p}\theta)_{\alpha}
\h
\not{p}_{\alpha\beta} = p_\nu \s^\nu_{\alpha\beta},\h
(\alpha = 1,2,..,16).
\ea
Here $(x^\nu, \theta^\alpha)$ are the superspace coordinates, 
$(p_\nu, p_{\theta\alpha})$ are their canonically conjugated 
momenta, $\theta^\alpha$ is a left Majorana-Weyl space-time spinor, 
and $\s^\mu$ are the ten dimensional Pauli matrices (our spinor 
conventions are as in \cite{PLB98}). The Hamiltonian (\ref{H}) is a 
generalization of the Hamiltonians for the bosonic null p-brane and 
for the $N=1$ Brink-Schwarz superparticle.

\section{\bf Solutions of the equations of motion}
\hspace{1cm}
The equations of motion which follow from the Hamiltonian $H_0$ are
$(\p_\tau = \p/\p\tau)$:
\ba\nl
(\pu-\mu^j\pj)x^\nu&=&2\mu^0p^\nu-(\mu\s^\nu\theta),\\
\label{em}
(\pu-\mu^j\pj)p_\nu&=&(\pj\mu^j)p_\nu,\\
\nl
(\pu-\mu^j\pj)\theta^\alpha&=&i\mu^\alpha,\\
\nl
(\pu-\mu^j\pj)p_{\theta\alpha}&=&(\pj\mu^j)p_{\theta\alpha}
+(\mu\not{p})_\alpha .
\ea
In (\ref{em}), one can consider $\mu^0$, $\mu^j$ and $\mu^\alpha$ 
as depending only on $\us=(\s^1,...,\s^p)$ but not on $\tau$ (this 
is a consequence from their equations of motion).

In the gauge when $\mu^0$, $\mu^j$ and $\mu^\alpha$ are constants,
the general solution of (\ref{em}) is
\ba
\nl
x^\nu(\tau,\us)&=&x^\nu(\uz)+\tau\bigl [2\mu^0p^\nu(\uz)-
(\mu\s^\nu\theta(\tau,\us))\bigr ], \\
\nl
&=&x^\nu(\uz)+\tau\bigl [2\mu^0p^\nu(\uz)-
(\mu\s^\nu\theta(\uz))\bigr ] \\
\label{gsp}
p_\nu(\tau,\us)&=&p_\nu(\uz) ,\\
\nl
\theta^\alpha(\tau,\us)&=&\theta^\alpha(\uz)+i\tau\mu^\alpha ,\\
\nl
p_{\theta\alpha}(\tau,\us)&=&p_{\theta\alpha}(\uz)+
\tau (\mu\s^\nu)_{\alpha} p_{\nu}(\uz) ,
\ea
where $x^\nu(\uz)$, $p_\nu(\uz)$, $\theta^\alpha(\uz)$ and 
$p_{\theta\alpha}(\uz)$ are arbitrary functions of their arguments
\ba\nl
z^j = \mu^j\tau+\s^j .
\ea

In the case of tensionless strings ($p=1$), one can write 
explicitly the general solution of the equations of motion in 
arbitrary gauge:  $\mu^0=\mu^0(\s)$, $\mu^1\equiv\mu=\mu(\s)$, 
$\mu^\alpha=\mu^\alpha(\s)$. This solution is given by
\ba
\nl
x^\nu(\tau,\s)&=&g^\nu(w)-2\int\limits_{}^\s\frac{\mu^0(s)}{\mu^2(s)}ds
f^\nu(w)+\int\limits_{}^\s\frac{\mu^\alpha(s)}{\mu(s)}ds
\bigl [\s^\nu\zeta(w)\bigr ]_\alpha \\
\nl
&-&i\int\limits_{}^\s ds_1\frac{(\mu\s^\nu)_\alpha(s_1)}{\mu(s_1)}
\int\limits_{}^{s_1}\frac{\mu^\alpha(s)}{\mu(s)}ds ,\\
\label{gs1}
p_\nu(\tau,\s)&=&\mu^{-1}(\s)f_\nu(w) ,\\
\nl
\theta^\alpha(\tau,\s)&=&\zeta^\alpha(w)-i\int\limits^\s
\frac{\mu^\alpha(s)}{\mu(s)}ds ,\\
\nl
p_{\theta\alpha}(\tau,\s)&=&\mu^{-1}(\s)\Biggl [h_\alpha(w)-
\int\limits^\s\frac{(\mu\s^\nu)_\alpha(s)}{\mu(s)}ds f_\nu(w)
\Biggr ] .
\ea
Here $g^\nu(w)$, $f_\nu(w)$, $\zeta^\alpha(w)$ and $h_\alpha(w)$ 
are arbitrary functions of the variable
\ba\nl
w = \tau + \int\limits^\s \frac{ds}{\mu(s)}
\ea

When $p=1$, the solution (\ref{gsp}) differs from (\ref{gs1}) by 
the choice of the particular solutions of the inhomogenious 
equations.  As for $z$ and $w$, one can write for example ($\mu^0$, 
$\mu$, $\mu^\alpha$ are now constants) 
\ba\nl 
p_\nu(\tau,\s)=\mu^{-1}f_\nu(\tau+\s/\mu)=\mu^{-1}f_\nu[\mu^{-1}
(\mu\tau+\s)]=p_\nu(z)
\ea
and analogously for the other arbitrary functions in the general 
solution of the equations of motion.

\section{\bf Lagrangian formulation}
\hspace{1cm}
Taking into account the equations of motion for $x^\nu$ and 
$\theta^\alpha$, one obtains the corresponding Lagrangian density
\ba
\nl
L=\frac{1}{4\mu^0}\Bigl [\bigl (\pu-\mu^j\pj\bigr )x+
i\theta\s\bigl (\pu-\mu^j\pj\bigr )\theta\Bigr ]^2 .
\ea
Indeed, one verifies that the equations of motion for the Lagrange 
multipliers $\mu^0$ and $\mu^j$ give the constraints $T_0$ and 
$T_j$. The remaining constraints follow from the definition of the 
momenta $p_{\theta\alpha}$.

To establish the invariances of the action, it is useful to rewrite 
$L$ in the form
\ba\nl
L=V^J V^K Y_J^\nu Y_{K\nu} \h,\h (J,K=0,1,...,p),
\ea
where
\ba\nl
V^J=\bigl (V^0,V^j\bigr )=\Biggl (-\frac{1}{2\sqrt{\mu^0}},
\frac{\mu^j}{2\sqrt{\mu^0}}\Biggr )
\ea
and
\ba\nl
Y_{J}^{\nu}=\p_{J} x^\nu+i(\theta\s^\nu\p_{J}\theta).
\ea
Then, the action
\ba\nl
S=\int d^{p+1}\xi V^J V^K Y_{J}^{\nu} Y_{K\nu}\h,\h
\xi^J = (\xi^0,\xi^j)=(\tau,\us),
\ea
has global super-Poincar$\grave{e}$ symmetry, 
local world-volume reparametrization and $\kappa$-invariances. 
Let us show that this is indeed the case. Before doing this, we 
note that actions of this type are first given in \cite{LST} for 
the case of tensionless superstring $(p=1)$ and in \cite{HLU} for 
the bosonic case $(N=0)$.

The global Poincar$\grave{e}$ invariance is obvious. Under global 
infinitesimal supersymmetry transformations, the fields
$x^\mu(\xi)$, $\theta^\alpha(\xi)$ and $V^J(\xi)$ transform as follows
\ba\nl
\delta_{\eta}\theta^{\alpha}=\eta^{\alpha}\h,\h
\delta_{\eta}x^{\mu}=i(\theta\s^{\mu}\delta_{\eta}\theta)\h,\h
\delta_{\eta}V^{J}=0 .
\ea
As a consequence, $\delta_{\eta}Y_{J}^{\nu}=0$ and hence
$\delta_{\eta}L=\delta_{\eta}S=0$ also.

To establish the invariance of the action under infinitesimal 
diffeomorphisms, we first write down the corresponding transformation law 
for the (r,s)-type tensor density of weight $a$ 
\ba\nl
\delta_{\varepsilon}T^{J_1...J_r}_{K_1...K_s}[a]&=&
L_{\varepsilon}T^{J_1...J_r}_{K_1...K_s}[a]=
\varepsilon^L\p_L T^{J_1...J_r}_{K_1...K_s}[a]\\
\label{r}
&+&
T^{J_1...J_r}_{KK_2...K_s}[a]\p_{K_1}\varepsilon^K+...+
T^{J_1...J_r}_{K_1...K_{s-1}K}[a]\p_{K_s}\varepsilon^K \\ \nl
&-&
T^{JJ_2...J_r}_{K_1...K_s}[a]\p_J\varepsilon^{J_1}-...-
T^{J_1...J_{r-1}J}_{K_1...K_s}[a]\p_J\varepsilon^{J_r} \\ \nl
&+&
aT^{J_1...J_r}_{K_1...K_s}[a]\p_L\varepsilon^L ,
\ea
where $L_\varepsilon$ is the Lie derivative along the vector field 
$\varepsilon$. Using (\ref{r}), one verifies that if $x^\mu(\xi)$, 
$\theta^\alpha(\xi)$ are world-volume scalars ($a=0$) and 
$V^J(\xi)$ is a world-volume (1,0)-type tensor density of weight 
$a=1/2$, then $Y_J^\nu$ is a (0,1)-type tensor, $Y_J^\nu Y_{K\nu}$
is a (0,2)-type tensor and $L$ is a scalar density of weight $a=1$.
So,
\ba\nl
\delta_{\varepsilon}S=\int d^{p+1}\xi\p_J\bigl (
\varepsilon^J L\bigr )
\ea
and the variation $\delta_{\varepsilon}S$ of the action vanishes 
under suitable boundary conditions.

Let us now check the kappa-invariance. We define the $\kappa$-
variations of $\theta^\alpha(\xi)$, $x^\nu(\xi)$ and $V^J(\xi)$
as follows:
\ba\nl
\delta_\kappa\theta^\alpha=i\bigl(\Gamma\kappa\bigr)^\alpha=
iV^J\bigl(\not{Y_J}\kappa\bigr)^\alpha,\\
\label{k}
\delta_\kappa x^\nu=-i(\theta\s^\nu\delta_\kappa\theta),\\
\nl
\delta_\kappa V^K=2V^K V^L (\p_L\theta\kappa).
\ea
Therefore, $\kappa^\alpha(\xi)$ is a left Majorana-Weyl 
space-time spinor and world-volume scalar density of weight 
$a=-1/2$. 

From (\ref{k}) we obtain:
\ba\nl
\delta_\kappa\bigl(Y_J^\nu Y_{K\nu}\bigr)=-2i\bigl[
\p_J\theta\not{Y_K}+\p_K\theta\not{Y_J}\bigr]
\delta_\kappa\theta
\ea
and
\ba\nl
\delta_\kappa L=2V^J Y_J^\nu Y_{K\nu}\bigl[\delta_\kappa V^K-
2V^K V^L(\p_L\theta\kappa)\bigr] = 0 .
\ea

The algebra of kappa-transformations closes only on the equations 
of motion, which can be written in the form:
\ba\nl
\p_J\bigl(V^JV^KY_{K\nu}\bigr)=0,\\
\label{eqm}
V^JV^K\bigl(\p_J\theta\not{Y_K}\bigr)_\alpha=0,\\
\nl
V^J Y_J^\nu Y_{K\nu}=0 .
\ea
As usual, an additional local bosonic world-volume symmetry is 
needed for its closure. In our case, the Lagrangian, and therefore 
the action, are invariant under the following transformations of 
the fields:
\ba\nl
\delta_\lambda\theta(\xi)=\lambda V^J\p_J\theta\h,\h
\delta_\lambda x^\nu(\xi)=-i(\theta\s^\nu\delta_\lambda\theta)\h,\h
\delta_\lambda V^J(\xi)=0 .
\ea
Now, checking the commutator of two kappa-transformations, we find:
\ba\nl
[\delta_{\kappa_1},\delta_{\kappa_2}]\theta^\alpha(\xi)&=&
\delta_\kappa\theta^\alpha(\xi)+
\mbox{terms $\propto$ eqs. of motion} , \\
\nl
[\delta_{\kappa_1},\delta_{\kappa_2}]x^\nu(\xi)&=&
(\delta_\kappa+\delta_\varepsilon+\delta_\lambda)x^\nu(\xi)+
\mbox{terms $\propto$ eqs. of motion} , \\
\nl
[\delta_{\kappa_1},\delta_{\kappa_2}]V^J(\xi)&=&
\delta_\varepsilon V^J(\xi)+
\mbox{terms $\propto$ eqs. of motion} .
\ea
Here $\kappa(\xi)$, $\lambda(\xi)$ and $\varepsilon(\xi)$ are given 
by the expressions:
\ba\nl
\kappa^\alpha=-2V^K[(\p_K\theta\kappa_1)\kappa_2^\alpha-
(\p_K\theta\kappa_2)\kappa_1^\alpha],\\
\nl
\lambda=4iV^K(\kappa_1\not{Y_K}\kappa_2)\h,\h
\varepsilon^J=-V^J\lambda .
\ea

We stress that 
\ba\nl
\Gamma_{\alpha\beta}=\bigl(V^J\not{Y_J}\bigr)_{\alpha\beta}
\ea
in (\ref{k}) has the following property on the equations of motion
\ba\nl
\Gamma^2 = 0 .
\ea
This means that the kappa-invariance of the action indeed halves 
the fermionic degrees of freedom as is needed.

Finally, we give the expression for world-volume stress-energy 
tensor
\ba
\label{t}
T^J_K=\bigl(2V^JY_K^\nu-\delta_K^JV^LY_L^\nu\bigr)V^MY_{M\nu}
\h,\h
Tr(T)=(1-p)L .
\ea
From (\ref{eqm}) and (\ref{t}) it is clear, that 
\ba
\label{con}
T^J_K=0
\ea
on the equations of motion. It is natural, because the equality 
(\ref{con}) is a consequence of $p+1$ of the constraints.

\section{\bf Conclusions}
\hspace{1cm}
In this letter we consider a model for tensionless (null) 
$p$-branes with $N=1$ global supersymmetry in 10-dimensional 
Minkowski space-time. We give an action for the model and show 
that it is reparametrization and kappa-invariant. 
As usual, the algebra of kappa-transformations closes only 
on-shell and it halves the fermionic degrees of freedom.
In proving the kappa-invariance, we do not use any specific ten 
dimensional properties of the spinors. Hence, the model is 
extendable classically to other space-time dimensions.
There exist also the possibility of its generalization to $N$
supersymmetries \cite{PLB99}. The properties of the model in 
nontrivial backgrounds are also under investigation \cite{BM}.

In this letter we also find some solutions of the classical equations 
of motion. In the case of null superstring ($p=1$), we obtain 
the general solution in arbitrary gauge.

\vspace*{1cm}

{\bf Acknowledgments}
\vspace*{.5cm}
\hspace{1cm}

The author would like to thank B. Dimitrov for careful reading of 
the manuscript.


\end{document}